\documentclass[aip,apl,twocolumn,preprintnumbers,amsmath,amssymb]{revtex4}

\usepackage{graphicx}
\usepackage[colorlinks=true,citecolor=blue]{hyperref}

\begin{document}
    %\linenumbers
    \title{The effect of photo-generated carriers on the spectral diffusion of a quantum dot coupled to a photonic crystal cavity}
    \author{Arka Majumdar}
    \email{arkam@stanford.edu}
    \author{Erik D. Kim}
    \author{Jelena Vu\v{c}kovi\'{c}}
    \affiliation{$^1$E.L.Ginzton Laboratory, Stanford University, Stanford, CA, $94305$\\}

    %\date{\today}
\begin{abstract}
We experimentally observe the effect of photo-generated carriers
on the spectral diffusion of a quantum dot (QD) coupled to a
photonic crystal (PC) cavity. In this system, spectral diffusion
arises in part from charge fluctuations on the etched surfaces of
the PC. We find that these fluctuations may be suppressed by
photo-generated carriers, leading to a reduction of the measured
QD linewidth by a factor of $\sim 2$ compared to the case where
the photo-generated carriers are not present. This result
demonstrates a possible means of countering the effects of
spectral diffusion in QD-PC cavity systems and thus may be useful
for quantum information applications where narrow QD linewidths
are desired.
\end{abstract}
\maketitle
\section{Introduction}
A single self-assembled quantum dot (QD) coupled to a high-quality
(Q) factor semiconductor optical cavity has emerged in recent
years as a promising system for developing on-chip quantum
technologies \cite{andrei_njp,article:eng07,article:ima99}.
Although this solid-state cavity quantum electrodynamic (CQED)
system behaves similarly to its atomic counterpart in a number of
ways, the unavoidable interaction of the QD with its solid-state
environment gives rise to phenomena specific to the solid-state
system. For instance, the presence of acoustic phonons even at
cryogenic temperatures has led to the observation of far
off-resonant coupling between a QD and a cavity, where the
resonant excitation results in the emission of a photon at the
cavity (QD) wavelength when the QD (cavity) is driven
\cite{article:majumdar09, article:michler09,hughes_mollow}. In
this case, the phonon-mediated off-resonant coupling may provide a
useful means of performing quantum state readout
\cite{majumdar_QD_splitting}.

Another important phenomenon arising from the solid state nature
of the QD-cavity system is the random trapping and untrapping of
charges in the vicinity of the QD due to defects, leading to
spectral diffusion of the QD optical transition energy
\cite{article_motional_narrowing,robinson_PRB,SD_CdSe_1,SD_CdSe_2}.
Spectral diffusion (or jitter) is generally considered detrimental
to the optical properties of QDs as it leads to the broadening of
optical emission lines
\cite{fluctuating_environment_PRB,Poizat_PRB,Poizat_nature}.
Studies of spectral diffusion in bulk self-assembled QD systems
have attributed its observation to defects provided by the highly
disordered QD wetting layer \cite{robinson_PRB,QD_defect_PRL}.
Additionally, it has been postulated that the proximity of etched
surfaces near the QD can lead to spectral diffusion due to defects
provided by the etched surface roughness \cite{imamoglu_sd_APL}.
Recent experimental studies of the linewidth of a QD
off-resonantly coupled to a photonic crystal (PC) cavity mode have
also observed broadening of QD linewidths that cannot simply be
attributed to a pure dephasing process \cite{article:majumdar10}.
This is not surprising considering that in QD-PC cavity systems
both the wetting layer and the presence of etched surfaces are
expected to contribute to spectral diffusion, particularly for QDs
close to the holes of the photonic lattice. Further, each
contribution is expected to manifest under different system
conditions. For instance, wetting layer defects predominantly trap
charges through the ionization of excitons generated in the
vicinity of the QD by above-band (AB) optical excitation. This
contribution to spectral diffusion has been shown to depend on the
strength of the AB laser and is absent at very low excitation
powers \cite{robinson_PRB}. For trap states provided by the rough
etched surfaces of the photonic crystal, fluctuating surface
charges are anticipated to contribute to spectral diffusion even
in the absence of AB optical excitation. It is unclear, however,
how this contribution to spectral diffusion may be affected by
above-band excitation, if it can be affected at all.

Here, we experimentally analyze the dependence of spectral
diffusion on the power of an above-band laser for a QD
off-resonantly coupled to a PC cavity. Specifically, we perform QD
linewidth measurements by scanning a narrow-bandwidth
continuous-wave (CW) laser across the QD resonance and observing
optical emission from the detuned cavity. In the absence of
above-band pumping, we observe a broadened QD linewidth and
attribute the broadening to fluctuating surface charges on the
nearby etched surfaces. Surprisingly, we find that this
contribution to spectral diffusion is suppressed in the presence
of above-band excitation. We attribute this suppression to the
filling of trap states in nearby etched surfaces by
photo-generated carriers, thereby drastically reducing charge
fluctuations. This in turn dramatically reduces the experimentally
observed QD linewidth. For low AB laser powers, spectral diffusion
becomes negligible and the QD linewidth exhibits the standard
power broadening as the CW laser power is increased. These results
demonstrate an optical excitation regime in which spectral
diffusion may be overcome and provide a simple means of countering
the degradation of the optical properties of QDs in PC structures.
This approach could thus prove useful in QD-PC cavity based
quantum technologies that benefit from narrower emitter
linewidths.

\section{QD Linewidth Broadening}
The effect of the spectral diffusion is reflected in the
measurement of the QD linewidth. In the linear response regime,
the QD lineshape is characterized by the QD susceptibility
\cite{Kubo_chapter}
\begin{equation}
\chi(t)\propto\mathcal{F}\left[e^{(i\omega_0-\frac{\gamma(\omega_0)}{2})t}\langle
e^{-i\int_0^td\tau\delta\omega(\tau)}\rangle\right]
\end{equation}
where $\mathcal{F}$ denotes the Fourier transform, $\omega_0$ is
the QD resonance frequency, $\gamma(\omega_0)$ is the radiative QD
linewidth and the statistical average is taken over all the
possible frequencies at different times $\tau$. The effect of
spectral diffusion is reflected in this statistical average. This
average depends on the underlying probability distribution of the
random environment. Assuming a Gaussian fluctuation, by Cumulant
expansion (and noting the odd moments are zero), we can write
\begin{equation}
\langle
e^{-i\int_0^td\tau\delta\omega(\tau)}\rangle=e^{-\frac{1}{2}\int_0^td\tau_1\int_0^td\tau_2\langle\delta\omega(\tau_1)\delta\omega(\tau_2)\rangle}
\end{equation}
Assuming an exponential correlation function (with variance
$\Gamma$ and correlation time $\tau_c$) of the form
\begin{equation}
\langle\delta\omega(\tau_1)\delta\omega{\tau_2}\rangle=\Gamma^2
e^{-\frac{|\tau_1-\tau_2|}{\tau_c}}
\end{equation}
we find that
\begin{equation}
\langle e^{-i\int_0^td\tau\delta\omega(\tau)}\rangle=e^{-\Phi(t)}
\end{equation}
with
\begin{equation}
\Phi(t)=\Gamma^2\tau_c^2[e^{-t/\tau_c}+t/\tau_c-1]
\end{equation}
Hence, we can model the effect of the fluctuating environment by
two quantities: the standard deviation $\Gamma$ and the
correlation time $\tau_c$. In the limit of long measurement time
($t>>\tau_c$), the linewidth of the QD is given by
($\gamma+\Gamma^2\tau_c$), where $\gamma$ comes from the
Lorentzian lineshape of the QD and $\Gamma^2\tau_c$ is the
contribution of the fluctuating environment. In this article, we
perform two separate sets of experiments to determine these
characteristic quantities of the fluctuating environment. From the
first set of experiments, we obtain a measure of the lifetime of
the charges in the traps and estimate the correlation time
$\tau_c$ to be approximately $100$ ns. In the second set of
experiments, we show how photo-generated carriers can mitigate the
effect of the spectral diffusion that is present in the absence of
any above-band excitation. From experimental results (see section
IV) we determine the broadening due to the fluctuating environment
and estimate the quantity $\Gamma^2\tau_c$ to be approximately
$8.7$ GHz, corresponding to $\Gamma=300$ MHz. We note that as
these values were obtained for two different QDs on the same
wafer, they provide only order-of-magnitude estimates.

\section{QD charging and modulation}
\begin{figure}
\centering
\includegraphics[bb = 0in 8.5in 3.25in 11in]{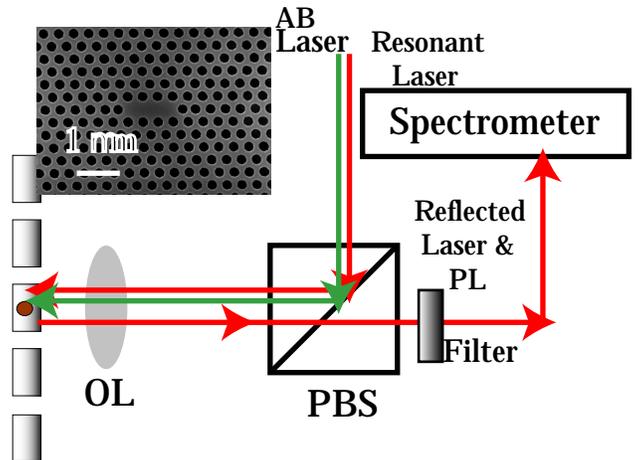}
\caption{(color online) The cross-polarized confocal microscopy
setup used to probe the coupled QD-cavity system. The inset shows
a scanning electron micrograph of the fabricated PC cavity. }
\label{fig_setup}
\end{figure}

All the experiments are performed in a helium-flow cryostat at
cryogenic temperatures ($\sim 30-55$ K) on self-assembled InAs QDs
embedded in a GaAs photonic crystal cavity \cite{article:eng07,
article:majumdar09}. The $160$nm GaAs membrane used to fabricate
the photonic crystal is grown by molecular beam epitaxy on top of
a GaAs $(100)$ wafer. The GaAs membrane sits on a $918$ nm thick
sacrificial layer of Al$_{0.8}$Ga$_{0.2}$As. Under the sacrificial
layer, a $10$-period distributed Bragg reflector, consisting of a
quarter-wave AlAs/GaAs stack, is used to increase the collection
into the objective lens \cite{article:eng07}. The photonic crystal
was fabricated using electron beam lithography, dry plasma
etching, and wet etching of the sacrificial layer in hydrofluoric
acid (6\%). The inset of Fig. \ref{fig_setup} shows the scanning
electron micrograph of a fabricated photonic crystal cavity. A
cross-polarized confocal microscopy setup is used to probe the
QDs, both in photoluminescence (PL) as well as under resonant
excitation as shown in Fig. \ref{fig_setup}
\cite{article:eng07,article:majumdar09}.

\begin{figure}
\centering
\includegraphics[bb = 0in 8.4in 3.25in 11in]{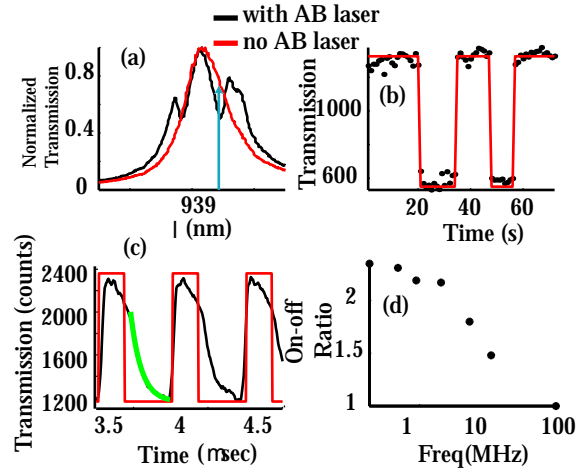}
\caption{(color online) (a) Effect of the above-band (AB) laser on
the transmission spectrum of a strongly coupled QD-cavity system.
The arrow shows the wavelength of the other (resonant) CW laser,
whose transmission is modulated by the AB laser in (b) and (c).
(b) The transmitted laser power in steady state when the AB laser
is turned on and off manually, at intervals of several $10$s. This
shows the DC characteristic of the modulation. (c) The temporal
dynamics of the QD-cavity system: the transmitted laser output as
a function of time (black plot). The shown modulation signal (red)
is used to modulate the above-band laser. From an exponential fit
(green) we estimate a time constant of $\sim 100$ ns. (d) The
on-off ratio of the transmitted laser as a function of the
modulation frequency. } \label{fig1}
\end{figure}

In the first set of experiments, we investigate the time scale on
which carriers generated by the above-band (AB) laser affect the
QD environment. These experiments are used to provide a measure of
correlation time $\tau_c$ associated with the fluctuating surface
charges. To perform these measurements, we first identify a
strongly coupled QD-cavity system in PL studies by observing the
anti-crossing of the QD and cavity resonances that occurs as the
temperature is varied. We then probe the system by measuring the
transmission of a resonant continuous wave laser through the
cavity. In the absence of the AB laser excitation, we observe a
Lorentzian line shape for the cavity transmission, indicating that
there is no QD coupled to the cavity despite the fact that
coupling is observed in PL studies. When the weak AB laser is
turned on, however, we observe a drastic change in the
transmission spectrum as shown in Fig. \ref{fig1}a. The spectrum
now appears to show two QDs strongly coupled to the cavity as
indicated by the presence of two dips in the transmission
spectrum. This dramatic change can be caused either due to
capturing of optically generated carriers by the QD
\cite{article:erik10}, or due to decreased spectral diffusion, and
thus reduction of the QD linewidth, as will be discussed later in
the paper. We note that in presence of the AB laser, the amount of
light coming out of the cavity due to PL is negligible compared to
the light transmitted through the cavity.

Next, we study the time dynamics of this resonance shift to obtain
a measure of the time scale on which the charge environment
surrounding the QD fluctuates. This is done by measuring the
transient response of the transmitted resonant laser power to
modulation of the AB laser. We note that this type of modulation
is different from the all-optical modulation reported in
\cite{dirk_opex_AM}, where the QD shifts continuously with
increasing AB laser power and the modulation is caused by the
screening of the built-in electric field by photo-generated
carriers. To obtain optimal modulation contrast we tune the CW
laser to the dip in the cavity transmission spectrum (shown by the
arrow in Figure \ref{fig1} a). Figure \ref{fig1} b shows the DC
modulation behavior when the above-band laser is manually turned
on and off at varying time intervals and the resonant laser
transmission is measured at steady-state. To measure temporal
dynamics, we apply a square-wave modulation to the above-band
laser and measure the CW laser transmission by a single photon
counter and pico-second time analyzer triggered by the modulated
above-band laser \cite{andrei_PRL_09}. We note that for this
experiment, the power of the above-band laser and the resonant CW
laser are, respectively, $15$ nW and $250$ nW as measured in front
of the objective lens of the confocal microscopy setup. Figure
\ref{fig1} c shows the modulated cavity transmission output with
time, for a modulation frequency of $2$ MHz. We observe an
exponential decay of the resonant laser transmission when the AB
laser is turned off. This is due to the fact that charge carriers
remain trapped for some amount of time even after the AB laser is
turned off. We fit the exponential decay and obtain a decay time
constant of $\sim 100$ ns, corresponding to a maximum modulation
speed of $\sim 10$ MHz. Finally Fig. \ref{fig1} d shows the
on-off-ratio of the transmitted CW laser as a function of the
modulation frequency of the above-band laser, which shows that the
maximum modulation speed is approximately on the order of $\sim
10$ MHz. From this experiment we estimate the time-scale of the
fluctuation of charges to be $\sim 100$ ns.

\section{QD Linewidth Broadening}
\begin{figure}
\centering
\includegraphics[bb = 0in 7.5in 3.25in 11in]{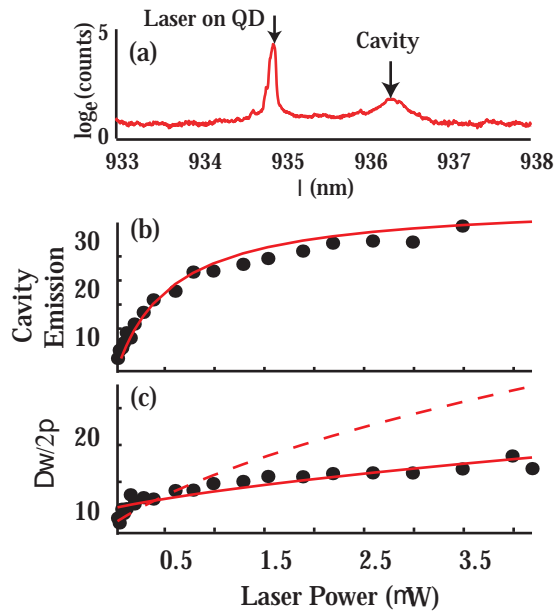}
\caption{(color online) QD spectroscopy by off-resonant dot-cavity
coupling: (a) The spectrum showing the laser resonantly driving
the QD and the emission from an off-resonant cavity. A logarithmic
scale is used to show both the strong laser and the weak cavity
emission. (b) Cavity emission as a function of the power of the
laser that is driving the QD resonantly. (c) QD linewidth
$\Delta\omega$ measured by monitoring the cavity emission by
scanning the laser across the QD. $\Delta\omega$ is plotted as a
function of the excitation laser power. We find that the standard
power broadening model over-estimates the measured QD linewidth
(the dashed line). However, the model provides better agreement
with experimental results when spectral diffusion is taken into
account (solid line).} \label{fig2}
\end{figure}

Having obtained a measure of the correlation time associated with
local charge fluctuations, we now turn to experimental
measurements of the broadening of QD linewidths by spectral
diffusion. For this set of experiments, we consider a system where
a QD is far off-resonantly coupled to a PC cavity mode. For this
particular system the dot is $1.4$ nm blue-detuned from the cavity
(the dot and cavity are, respectively, at $934.9$ and $936.3$ nm;
see Figure \ref{fig2} a). QD linewidths are obtained by measuring
the off-resonant cavity emission as the CW laser is scanned
through the QD resonance, similar to experiments reported in Ref.
\cite{article:majumdar10}. We observe saturation of the cavity
emission and power broadening of the QD with increasing excitation
laser power. We emphasize that in the experimental data shown in
Fig. \ref{fig2}, the additional above-band excitation is not used,
and the QD is driven only by the resonant laser. By simultaneously
fitting the cavity emission with the model
$I_{sat}\tilde{P}/(1+\tilde{P})$ (Figure \ref{fig2} b) and the QD
linewidth with the model $\Delta\omega_0\sqrt{1+\tilde{P}}$
(Figure \ref{fig2} c) \cite{article:shih09}, we find that the
observed QD linewidth is significantly smaller than what is
expected from ordinary power broadening of a two level system (the
dashed line in Figure \ref{fig2} c). Here, $I_{sat}$ is the
saturated cavity emission intensity, $\Delta\omega_0$ is the QD
linewidth at the limit of zero excitation power and $\tilde{P}$ is
the proportional to the laser intensity at the position of the QD.
Simple theoretical analysis shows that the QD linewidth exhibits
power broadening similar to that of a two-level system
\cite{majumdar_phonon_11}, thus showing that the cavity read-out
does not affect the QD linewidth significantly. This unusual
broadening behavior cannot be explained by an increasing pure
dephasing rate with increasing laser power or excitation induced
dephasing \cite{EID_ramsay}, since those effects would lead to a
broader linewidth than that predicted by standard power
broadening. In fact, the nature of the broadening indicates a
power-independent broadening mechanism.

We thus consider spectral diffusion as a means of modeling this
additional power-independent broadening. We note that spectral
diffusion is indeed power independent when the laser is resonant
with the QD. The effect of spectral diffusion is modeled as a
Voigt lineshape, which is a convolution of a Lorentzian lineshape
(the actual QD linewidth $\Delta\omega_L$) and a Gaussian
lineshape (the spectral fluctuations with a linewidth of
$\Delta\omega_D$). The full width half maximum $\Delta\omega_V$ of
a Voigt line-shape is given by \cite{voigt_profile}:
\begin{equation}
\Delta\omega_V\approx
A\Delta\omega_L+\sqrt{(1-A)^2\Delta\omega_L^2+\Delta\omega_D^2}
\end{equation}
where the parameter $A=0.5346$ is empirically determined. We use
this model to fit the power broadened linewidth, where
$\Delta\omega_L=\Delta\omega_0\sqrt{1+\tilde{P}}$. The model fits
the experimental data very well, with parameters
$\Delta\omega_0/2\pi=4.7$ GHz and $\Delta\omega_D/2\pi=8.7$ GHz.
These results support the theory that spectral diffusion is caused
by the empty traps present on the etched surfaces of the photonic
crystal. These traps can be randomly charged and discharged,
resulting in a fluctuating charge environment that can modify the
QD resonance frequency by random DC Stark shifts. As we perform
time-averaged measurements on time scales much longer than the
correlation time estimated in the preceding set of experiments,
the measured QD linewidth is a statistical mixture of the QD
resonances at different resonance frequencies.

\begin{figure}
\centering
\includegraphics[bb = 0in 8.5in 3.25in 11in]{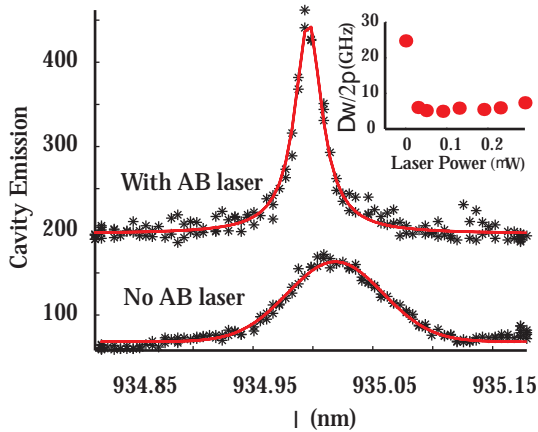}
\caption{(color online) Effect of the AB laser on the QD spectrum
as measured though off-resonant cavity emission. In the absence of
the AB laser, we observe a broad QD linewidth. However, in the
presence of this laser, a significant linewidth narrowing is
observed. The inset shows the QD linewidth measured as a function
of the AB laser power.} \label{fig4}
\end{figure}

We now repeat these experiments in the presence of an above-band
laser. In this system, although the QD transition energy is not
shifted in the presence of the PL laser, we observe a significant
reduction in the measured QD linewidth (by a factor of $\sim 2$)
and an increase in the off-resonant cavity emission (Figure
\ref{fig4}). The resonant laser power is kept at the same value
($\sim 100$ nW) as used in the measurements of Fig. \ref{fig2}.
The increased background in the cavity emission is caused by the
PL generated by the AB laser. We find that the QD exhibits a
Lorentzian line-shape in presence of the AB laser, but shows a
nearly Gaussian lineshape in the absence of the AB laser. We
attribute this difference to the fact that one of the effects of
the AB laser is to create carriers that subsequently fill all the
empty traps near the QD, thereby reducing spectral fluctuations.
The dependence of the QD linewidth as a function of above-band
laser power is shown in the inset of Fig. \ref{fig4}. With
increasing AB laser power, the PL adds noise to the off-resonant
cavity emission, making linewidth measurements difficult. However,
we do observe a slight increase in the QD linewidth with
increasing AB laser power, consistent with previous studies
\cite{robinson_PRB}. We note that an AB laser can cause multiple
effects such as an abrupt change in the QD resonance frequency
\cite{article:erik10}, or saturation of the QD. Hence this type of
linewidth narrowing in the presence of the AB laser is not
observed for all the QDs. However, the additional
power-independent broadening is present in the absence of
above-band pumping for all QDs studied.

\begin{figure}
\centering
\includegraphics[bb = 0in 8.5in 3.25in 11in]{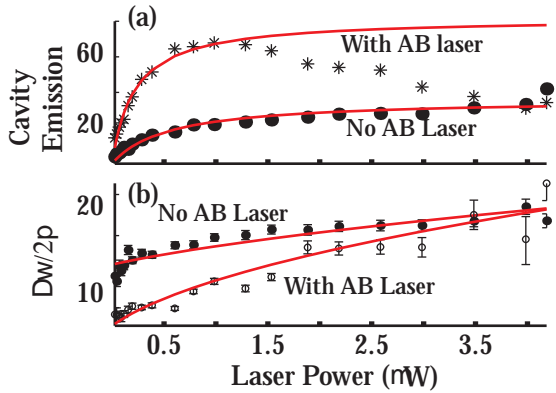}
\caption{(color online) (a) Off-resonant cavity emission as a
function of the laser power driving the QD resonantly for two
cases: with and without the additional, weak above band laser. (b)
QD linewidth as a function of the resonant driving laser power. A
much narrower linewidth is observed in the presence of the
above-band laser at lower excitation powers. Also, the QD
linewidth follows a simple power broadening with the excitation
laser power when the above-band laser is present. In the absence
of the above-band laser, we observe an additional power
independent broadening, which can be explained by spectral
diffusion of the QD.} \label{fig3}
\end{figure}

For this particular QD, we perform a power dependent study of the
QD linewidth and cavity emission in the presence of the AB laser.
The AB laser power is kept constant throughout experiments. The
data in the absence of the AB laser is the same as shown in Figure
\ref{fig2}. Figure \ref{fig3} a shows the off-resonant cavity
emission intensity as a function of the excitation laser power
both in the presence and absence of the AB laser. In the presence
of the AB laser, we observe an increased cavity emission and also
a saturation trend at lower laser power (shown by the fit).
However, with increasing excitation laser power, the cavity
emission decreases. This decrease in the cavity emission may be
due to the presence of other QD excitonic states that the
resonantly pumped QD is coupled to. The fact that those states are
present only with the AB excitation may be an indication that
these other excitonic states are multiply charged states arising
from the capture of multiple photo-generated carriers
\cite{karrai_QD_complex}. Figure \ref{fig3} b shows the measured
QD linewidth as a function of the excitation laser power both in
the presence and absence of the AB laser. We find that the QD
linewidth is narrower with AB laser on, at lower laser excitation
power. In fact, the QD linewidth in that case follows ordinary
power broadening (as shown by the solid line fit), with
$\Delta\omega_0/2\pi=4.7$ GHz, the same as the Lorentzian
linewidth obtained from the fit to the linewidth data with
spectral diffusion.

In summary, we have presented experimental studies showing the
effect of the photo-generated carriers on spectral diffusion in
the QD-cavity QED system. We first showed that the transmission
spectrum of the strongly coupled QD-cavity system can be modified
with a very low power of the additional above-band laser. We also
show the significant effect of spectral diffusion in resonant QD
spectroscopy, when the QD is embedded in a photonic crystal
cavity. This spectral diffusion can be mitigated by
photo-generated carriers that serve to fill the trap states
provided by the etched surfaces of the PC. This significant
reduction of the QD linewidth by AB excitation is an attractive
prospect for potential quantum information applications employing
QDs in PC structures. For those applications benefiting from
narrow QD linewidths, the results presented here demonstrate a
relatively simple means of improving the optical quality of QDs
and thus system performance.

The authors acknowledge financial support provided by the NSF,
ARO, and ONR. E.K. acknowledges support from the IC Postdoctoral
Research Fellowship. The authors acknowledge Dr. Pierre Petroff
and Dr. Hyochul Kim for providing the QD sample.
\bibliography{NRDC_bibl}
\end{document}